\documentclass[universe,article,accept,pdftex,moreauthors]{Definitions/mdpi} 

\firstpage{1} 
\pubvolume{1}
\issuenum{1}
\articlenumber{0}
\pubyear{2024}
\copyrightyear{2024}
\externaleditor{\textls[-25]{Academic Editor: Firstname Lastname}}
\datereceived{5 May 2024} 
\daterevised{1 August 2024} 
\dateaccepted{20 August 2024} 
\datepublished{ } 
\hreflink{https://doi.org/} 

\Title{Source Count Distribution of Fermi LAT Gamma-Ray Blazars Using Novel Nonparametric Methods} 

\TitleCitation{Source Count Distribution of Fermi LAT Gamma-Ray Blazars Using Novel Nonparametric Methods}

\Author{Xuhang Yin $^{1,2}$ and Houdun Zeng $^{1,*}$\orcidA{}}

\AuthorNames{Xuhang Yin and Houdun Zeng}

\AuthorCitation{Yin, X.; Zeng, H.}

\address{%
$^{1}$ \quad Key Laboratory of Dark Matter and Space Astronomy, Purple Mountain Observatory, Chinese Academy of Sciences, Nanjing 210023, China; {yinxh95@163.com}\\
$^{2}$ \quad Department of Astronomy, Yunnan University, Kunming 650091, China}

\corres{Correspondence: zhd@pmo.ac.cn}

\abstract{We utilized a sample from the Fermi-LAT 14-year Source Catalog by adjusting the flux detection threshold, enabling us to derive the intrinsic source count distribution $dN/dF_{25}$ of extragalactic blazars using nonparametric, unbinned methods developed by Efron and Petrosian and Lynden-Bell. Subsequently, we evaluated the contribution of blazars to the extragalactic gamma-ray background. Our findings are summarized as follows:
(1) There is no significant correlation between flux and spectral index values among blazars and their subclasses FSRQs and BL Lacs. (2) The intrinsic differential distributions of flux values exhibit a broken-power-law form, with parameters that closely match previous findings. 
The intrinsic photon index distributions are well described by a Gaussian form for FSRQs and BL Lacs individually, while a dual-Gaussian model provides a more appropriate fit for blazars as a whole.
(3) Blazars contribute 34.5\% to the extragalactic gamma-ray background and 16.8\% to the extragalactic diffuse gamma-ray background. When examined separately, FSRQs and BL Lacs contribute 19.6\% and 13\% to the extragalactic gamma-ray background, respectively.}


\begin{document}




\section{Introduction} \label{sec:intro}

The extragalactic gamma-ray background (EGB) refers to diffuse gamma-ray radiation originating beyond our Milky Way galaxy. This background consists of contributions from various astrophysical sources, such as active galactic nuclei (AGNs), star-forming galaxies, and potentially dark matter annihilation or decay. Understanding the composition and origin of the EGB is crucial for advancing our knowledge of high-energy astrophysics and cosmology. A prominent category of astrophysical objects, blazars, which encompass both flat-spectrum radio quasars (FSRQs) and BL Lacertae objects, represent a dominant class of radio-loud AGNs detected by the Large Area Telescope (LAT) of the Fermi Gamma-Ray Space Telescope (see, e.g.,
\,\cite{2010ApJS..188..405A,2020ApJS..247...33A,2020arXiv200511208B}). These blazars are highly variable and luminous, emitting intense gamma-ray radiation due to relativistic jets nearly aligned with Earth's line of sight (see, e.g., \cite{1995PASP..107..803U,2009ApJ...692...32D,2016ApJS..226...20F,2017ApJS..228....1Z,2018ApJS..235...39C}).

Generally, the contribution of blazers to the EGB is typically estimated using the luminosity function and its evolution with various parameters, as seen in, e.g., \cite{2012ApJ...751..108A,Ajello_2013,2014ApJ...780..161D,2014ApJ...786..129D,
2012ApJ...749..151Z,2013MNRAS.431..997Z,2019MNRAS.490..758Q}. 
However, determining these distributions often requires variables such as redshift (z), which are not readily available for most blazars, especially BL Lacs objects. Conversely, blazars' contribution to the EGB can be directly estimated if the source count distributions of gamma-ray emitters (log$S$-log$N$) are provided, as seen in, e.g., \citep{2010ApJ...720..435A,2012ApJ...753...45S,2021ApJ...913..120Z,2016PhRvL.116o1105A}. Examining the population statistics of gamma-ray sources can provide valuable insights into the cumulative emissions from these objects and their overall impact on the EGB. 

Determining the distributions of flux and photon spectral index values is complex due to the limitations of Fermi-LAT in relation to its threshold flux values. 
 For instance, blazars with harder spectra are more easily detected at lower flux levels. Additionally, each detected source has a different threshold flux, determined by a detection test statistic (TS), because the background flux varies with the position on the sky \cite{2010ApJ...715..429A}.
Observational biases have been analyzed and mitigated using Monte Carlo simulations (see, e.g., \cite{2010ApJ...720..435A,2020ApJ...896....6M}). Moreover, Refs.
 \,\cite{2012ApJ...753...45S,2021ApJ...913..120Z} employed a method developed by Efron and Petrosian (EP) \cite{1992ApJ...399..345E} and Lynden-Bell \citep{1971MNRAS.155...95L} to directly determine the distributions from observational data, accounting for complex observational selection biases, initially identifying the intrinsic correlations between variables and subsequently establishing the univariate distribution of each variable. The advantage of these methods lies in their avoidance of ad hoc assumptions and their direct extraction of the mono-variate distributions from data in a nonparametric manner, without resorting to binning. Nonparametric methods have successfully contributed to the study of AGNs (e.g.,~\cite{1999ApJ...518...32M,2011ApJ...743..104S,2012ApJ...753...45S,2013ApJ...764...43S,2014ApJ...786..109S,2021ApJ...913..120Z}) and gamma-ray bursts (e.g., \cite{2015ApJS..218...13Y,2015ApJ...806...44P,2016A&A...587A..40P,2017ApJ...850..161T}).

In this paper, we apply nonparametric methods to the truncated data detected by Fermi-LAT in order to ascertain the intrinsic correlations between photon flux and the photon index, followed by analyzing their intrinsic distributions. In contrast to Refs. \cite{2012ApJ...753...45S,2021ApJ...913..120Z}, we have employed a novel sample selection method to procure suitable samples and assess the detection efficiency of blazars. We note that the study in \cite{2021ApJ...913..120Z} focuses on variable luminosity and redshift, analyzing the evolution of the gamma-ray and optical luminosity functions of blazars. In contrast, here, we do not deal with cosmological evolution, and instead focus solely on a detailed analysis of the intrinsic distribution of flux and the photon index.  
This paper is organized as follows: Section~\ref{sec:sample} details our selection of an available sample of data from the Fermi-LAT extragalactic catalog. Section~\ref{sec3} focuses on analyzing these data to obtain the intrinsic distribution of flux and photon spectral index values and subsequently calculating the contribution of blazars to the EGB. Section~\ref{sec4} provides a brief summary and discussion.

\section{Sample} \label{sec:sample}
The fourth Fermi-LAT AGN catalog, 4LAC \citep{2020ApJ...892..105A}, comprises 2863 AGNs of various types detected by Fermi-LAT operating between 4 August  2008 and 2 August 2016, situated at galactic latitude $|b| \geq 10^o$. After excluding entries in 4LAC not associated with known AGNs, those with multiple counterparts, or flagged for other reasons in the 4FGL FITS file\endnote{\url{https://fermi.gsfc.nasa.gov/ssc/data/access/lat/8yr_catalog/gll_psc_ v20.fit}, (accessed on 6 August 2020).},\,a clean sample of 2614 sources emerged, including 591 FSRQs, 1027 BL Lacs, 941 blazars of unknown type (BCUs), and 55 non-blazar AGNs. Analysis of the energy flux distributions for high-latitude sources in 4LAC (Figure 4 \,of \cite{2020ApJ...892..105A}) has suggested that the detection threshold approximates a truncation line where the LAT energy flux threshold $S_{25,\rm lim}$ rises with increasing spectral index $\Gamma$. This line was thus chosen for use in this study because the updated sample from the Fermi group does not provide the expected detection threshold.  

\begin{figure}[H]
\includegraphics[width=7.0 cm]{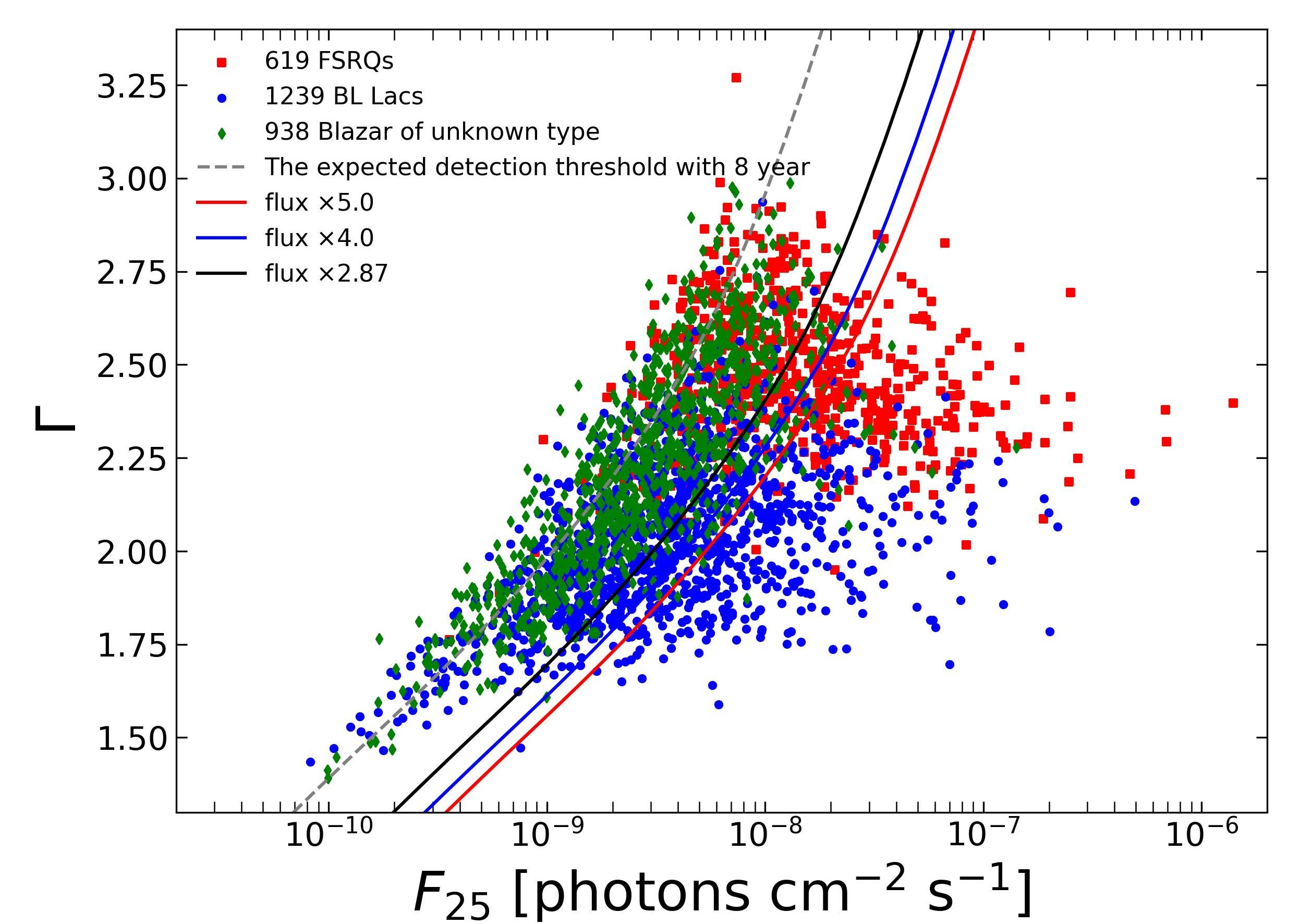}
\includegraphics[width=7.0 cm]{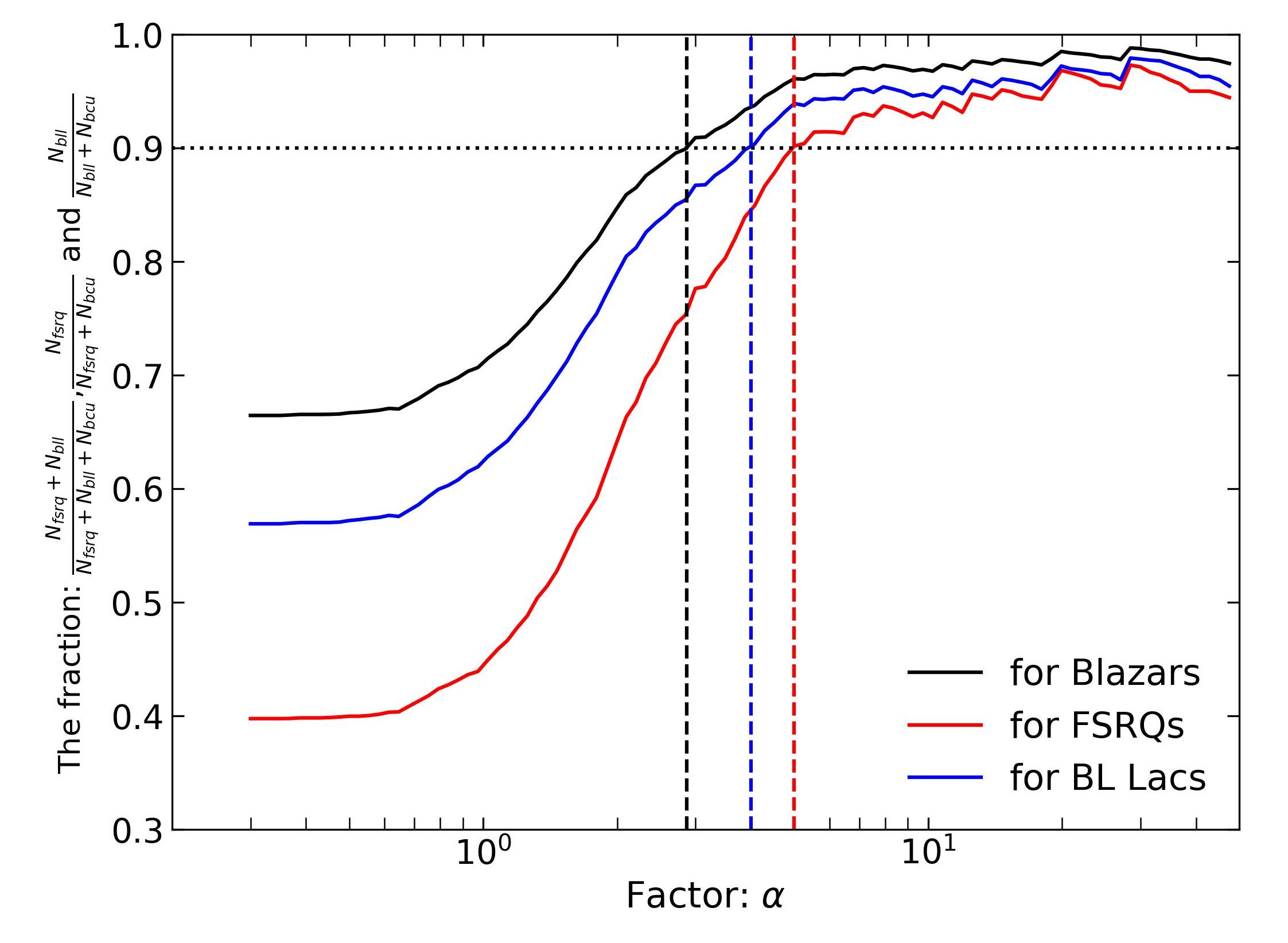}        
\caption{\textbf{Left}: Scatter diagram of the photon flux $F_{25}$ vs. the power-law index $\Gamma$. The truncation curve (dashed line) shows the expected detection threshold with 8-year Fermi-LAT data ($F_{25, \rm lim}$ vs. $\Gamma_{\rm lim}$), which reflects a larger dependence of the detection threshold on the power-law index (from Figure 16 of \cite{2020ApJS..247...33A}). \textbf{Right}: The fraction of source $\frac{N_{\rm fsrq}+N_{\rm bll}}{N_{\rm fsrq}+N_{\rm bll}+N_{\rm Bcu}}$, $\frac{N_{\rm fsrq}}{N_{\rm fsrq}+N_{\rm Bcu}}$, $\frac{N_{\rm bll}}{N_{\rm bll}+N_{\rm Bcu}}$ vs. the shift factor $\alpha$. The black, red, and blue lines show the fraction of blazar, FSRQ, and BL Lac objects with the shift factor, respectively. When the shift factor $\alpha=2.87$, the fraction of blazars reaches 0.9, while the shift factors of FSRQs and BL Lacs need to be equal to 5.0 and 4.0.
\label{fig:fgGam}}
\end{figure}

An updated version (4FGL$-$DR4, for Data Release 4) of the fourth full catalog of LAT sources, based on 14 years of survey data in the 50 MeV--1 TeV energy range, is now available online \endnote{\url{https://fermi.gsfc.nasa.gov/ssc/data/access/lat/14yr_catalog/}, (accessed on 20 December 2023).}. 
Initially, a sample was selected from the LAT 14-year Source Catalog, available as a FITS file \endnote{\url{https://fermi.gsfc.nasa.gov/ssc/data/access/lat/14yr_catalog/gll_psc_v33.fit}, (accessed on 20 December 2023).}, under the conditions of a source significance over 4.3 $\sigma$ (TS $>25$, fitted with a power-law function) and analysis flags equal to 0, resulting in 2796 blazars, including 619 FSRQs, 1239 BL Lacs, and 938 BCUs. For our analysis, we required the photon flux ranging from 100 MeV to 100 GeV and the photon index. Although the photon flux is not provided in the sample, it can be derived using the following expression: 

\begin{eqnarray}
F_{25}= S_{25} /E_1\times 
 \left\{
\begin{array}{lcl}
 \frac{\Gamma-2}{\Gamma-1}\times  \frac{1-10^{-3(\Gamma-1)}}{1-10^{-3(\Gamma-2)}} &
& {\rm (if} \,\,\, \Gamma \neq 2.0) \\
\frac{(1-10^{-3})}{\ln(10^3)} &
& {\rm (if} \,\,\, \Gamma = 2.0)\;,
\end{array}
\right.
\label{eq:S-F}
\end{eqnarray}
where $E_{1}=1.602 \times 10^{-4}$ erg (corresponding to 100 MeV) represents the lower energy limit. $F_{25}$ and $S_{25}$\endnote{The subscript 2 in $F_{25}$ and $S_{25}$ represents 100 MeV and 5 represents 100 GeV.}, respectively, denote the photon flux and the energy flux from 100 MeV to 100~GeV, and  $\Gamma$ represents the absolute value of the photon index.
The scatter diagram in the left panel of Figure \ref{fig:fgGam} illustrates the photon flux $F_{25}$ versus the power-law index $\Gamma$,  with the dashed line representing the expected detection threshold with 8-year Fermi-LAT data ($F_{25, \rm lim}$ vs. $\Gamma_{\rm lim}$). This curve underscores the detection threshold's substantial dependence on the power-law index (from Figure 16 of \cite{2020ApJS..247...33A}). 
Due to the significant number of BCUs, they may be classified as either FSRQs or BL Lacs, leading to substantial classification incompleteness in the samples of both types; i.e., causing the classification fraction of source $\frac{N_i}{N_i+N_{\rm Bcu}}$, where $i$ represents FSRQs or BL Lacs, to be low. To mitigate this incompleteness of the FSRQ and BL Lac sample, we introduced a simple translation of the detection threshold by constructing a shift factor $\alpha$, such that $\acute{F}_{25, \rm lim}$ = $\alpha F_{25, \rm lim}$ for a fixed $\Gamma_{\rm lim}$. All sources with $F_{25}>\acute{F}_{25, \rm lim}$ ( or $S_{25}>\acute{S}_{25, \rm lim}$) were selected until the  classification fraction reached 0.9 by varying the value of $\alpha$. Consequently, we obtained two samples: 234 FSRQs corresponding to $\alpha=5.0$ and 466 BL Lacs corresponding to $\alpha=4.0$. The FSRQ sample is designed to include 90\% of sources identified as FSRQs by excluding known BL Lacs, while the BL Lac sample is designed to encompass 90\% of the sources identified as BL Lacs by removing known FSRQs. For the blazar sample, we used a more conservative limit with $\alpha=2.87$ and a fraction $\frac{N_{\rm fsrq}+N_{\rm bll}}{N_{\rm fsrq}+N_{\rm bll}+N_{\rm Bcu}}$ = 0.9. This resulted in the selection of 1003 blazars, ensuring that 90\% of all sources could be identified as either FSRQs or BL Lacs. Therefore, we utilize three different samples in the following analysis. The right panel of Figure \ref{fig:fgGam} shows the relationship between the fraction of sources and the translation factor.

\section{Analysis and Results} \label{sec3}
The aim of this study is to determine the intrinsic source count distribution of blazars and evaluate their contribution to the EGB.
The previous section provides us with a description of the sample, which includes bi-variate truncated data. This requires investigating the independence between the photon flux and photon index. If these two variables are independent, their distributions, represented as $G(F_{25},\Gamma)$, can be separated into two distinct distributions: $\psi(F_{25})$ and $h(\Gamma)$. However, in scenarios like gamma-ray bursts, there may be a persistent correlation between luminosity and the photon index due to the intrinsic relationship between flux and photon index values, even after mitigating cosmic influences \citep{2004ApJ...609..935Y,2000ApJ...534..227L}. Therefore, to isolate the intrinsic distributions within our sample, it is crucial to eliminate the correlation between flux and the photon index.

\subsection{Correlations} \label{sec3.1}
\begin{figure}[H]
\includegraphics[width=6.5 cm]{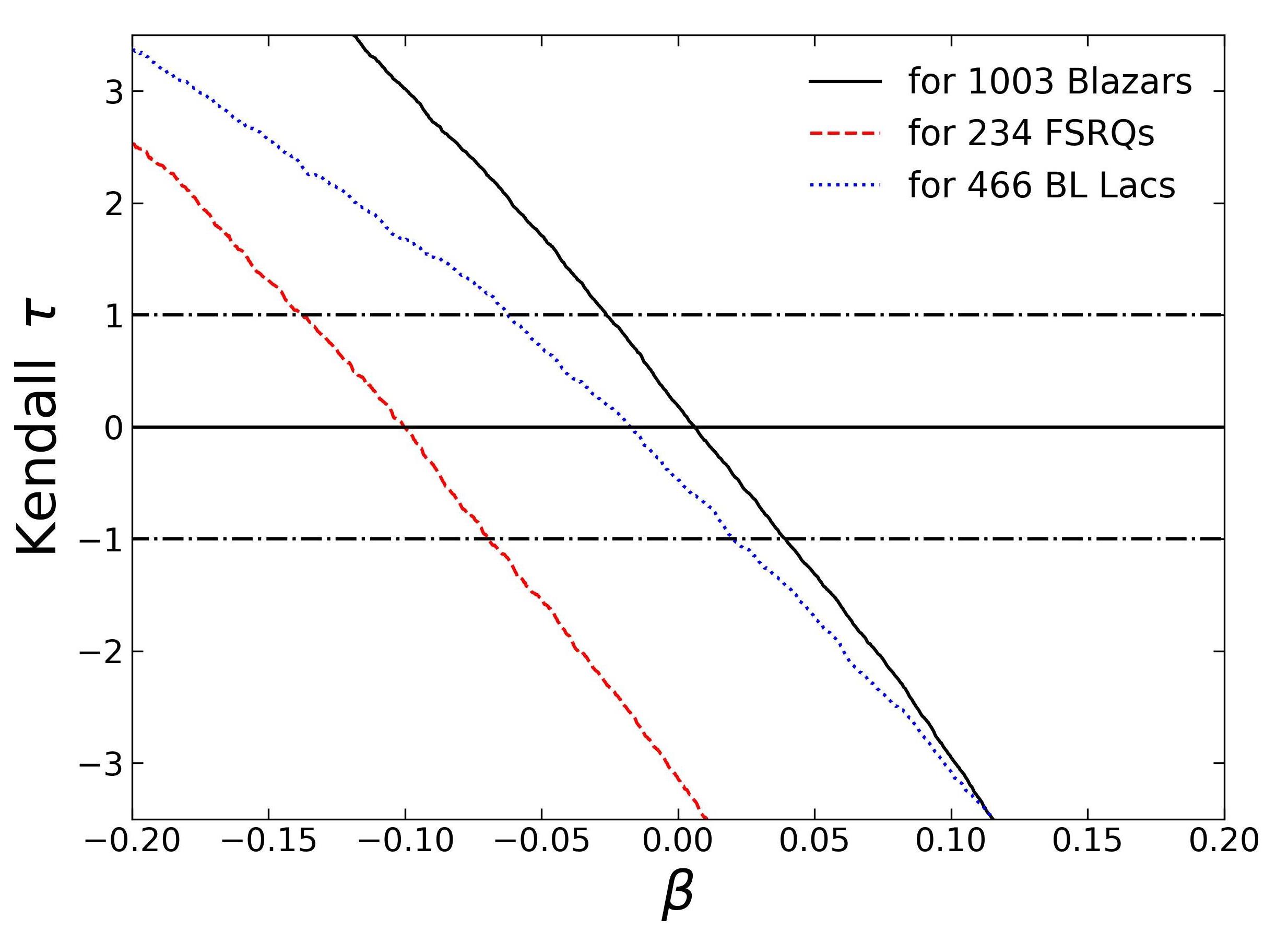}
\includegraphics[width=6.5 cm]{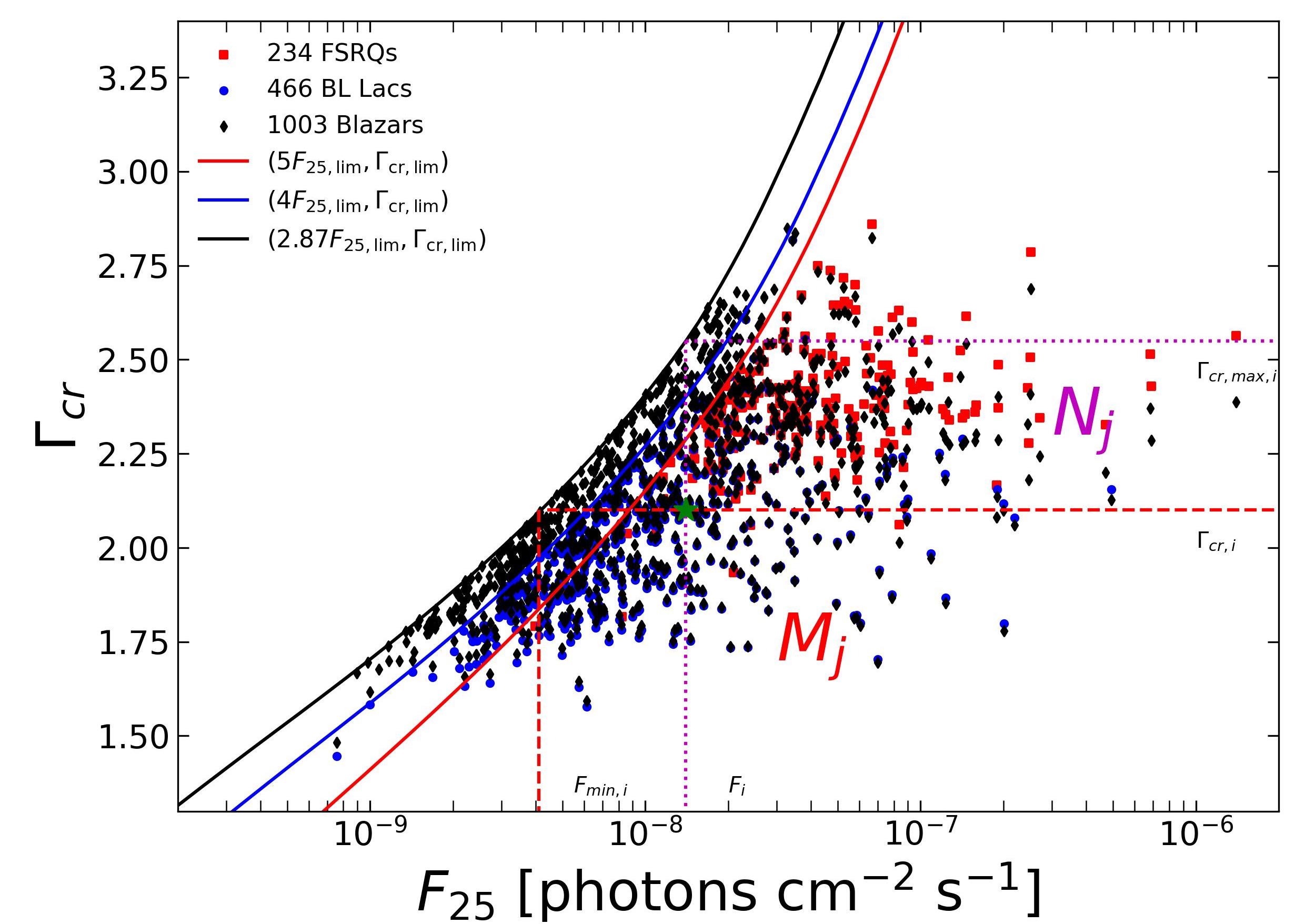}
\caption{\textbf{Left}: The value of the test statistic $\tau$ as a function of $\beta$ for blazars, FSRQs, and BL Lacs, which indicates a correlation between the photon index and flux. The values of $\beta$ for which $\tau=0$ and
$\pm 1$ give the best value and 1-sigma range, respectively. A weak correlation or no correlation  are indicated with $\beta_{\rm blazars}=0.006 \pm 0.033$ for blazars, $\beta_{\rm fsrq}=-0.100 ^{+0.031}_{-0.038}$ for FSRQs, and $\beta_{\rm bll}=-0.017^{+0.037}_{-0.045}$ for BL Lacs. \textbf{Right}: Scatter diagram of photon flux $F_{25}$ vs the correlation-reduced photon index $\Gamma_{\rm cr}$. The black, red, and blue curves show the flux threshold limit $\alpha F_{25}$ vs. $\Gamma_{\rm cr}$ (the truncation curve) for blazars, FSRQs, and BL Lacs, respectively, with the different shift factor. The two boxes enclosed by the dashed (red) and dotted (purple) lines, respectively, represent two associated sets ($M_j$ and $N_j$) of the source marked by the green star, as defined in Equation (6). 
\label{fig:tau_alpha}}
\end{figure}

Here, we employ the Efron–Petrosian method to assess the correlation between two variables. This method presents a variation of the Kendall Tau statistic test tailored for truncated data, utilizing a test statistic to examine the independence of two variables within a dataset, as follows:
\begin{equation}
\tau=\frac{\Sigma_i(R_i-E_i)}{\sqrt{\Sigma_iV_i}},    
\end{equation}
where $R_i$ is the normalized ranks of the sources within their {\it associated sets}, defined as sources with $\Gamma_j<\Gamma_i$ and $F_j>F_{i,{\rm lim}}$, for ranking in $F$.
Further details on this test can be found \linebreak in~\cite{1992ApJ...399..345E,1999ApJ...518...32M,2014ApJ...786..109S,2015ApJ...806...44P}. Under the null hypothesis of independence, the expected ranks and variances are $E_i=1/2$  and $V_i=1/12$. Our findings indicate that for both subsamples, $\tau(k=0)>5$, rejecting the null hypothesis at a significance level exceeding $5\sigma$. To investigate the correlation between photon flux and photon index, we follow the methodology delineated in \cite{2012ApJ...753...45S}. In this approach, a new parameter, termed the correlation-reduced photon index ($\Gamma_{\rm cr}$), is introduced, defined as follows:
\begin{equation}
    \Gamma_{\rm cr}=\Gamma-\beta \times {\rm log_{10}}\left(\frac{F_{25}}{F_{0}}\right).
    \label{eq: Gamma_cr}
\end{equation}
This transformation is essentially a simple coordinate rotation. The parameter $\beta$ is empirically determined such that $F_{25}$ and $\Gamma_{\rm cr}$  are independent {\bf ($\tau=0$)}, allowing the combined distribution to be expressed as
$G(F_{25},\Gamma)=\psi(F_{25}) \times \hat{h}(\Gamma_{\rm cr})$.
Once the photon flux distributions are determined, the intrinsic distribution of the photon index ($h(\Gamma)$) is obtained through integration over flux:
\begin{equation}
h(\Gamma)=\int_{F_{25}}^{\infty} \psi(F_{25}) \hat{h}\left(\Gamma-\beta \times {\rm log_{10}}\left(\frac{F_{25}}{F_{0}}\right)\right) dF_{25}.
\label{eq:hGam}
\end{equation}
Here, $F_{0}=3.0 \times 10^{-8}$ photon cm$^{-2}$ s$^{-1}$ serves as a fiducial flux, with its specific value being noncritical. It is essential to transform the truncation curve $\Gamma_{\rm lim}$ in the $F_{25} - \Gamma$ plane using the same parameter $\beta$,
$\Gamma_{\rm cr, lim}=\Gamma_{\rm lim}-\beta \times {\rm log_{10}}(\frac{F_{25}}{F_{0}})$.
The left panel of Figure \ref{fig:tau_alpha} shows the variation in $\tau$ with the correlation parameter $\beta$, displaying the best-fit values and 1-sigma ranges of $\beta_{\rm blazars}=0.006 \pm 0.033$ for blazars, $\beta_{\rm fsrq}=-0.100 ^{+0.031}_{-0.038}$ for FSRQs, and $\beta_{\rm bll}=-0.017^{+0.037}_{-0.045}$ for BL Lacs. The results indicate a weak correlation or no correlation for all sources, consistent with the findings of \citet{2012ApJ...753...45S}. The right panel of Figure \ref{fig:tau_alpha} shows a scatter diagram of the photon flux $F_{25}$ vs. the correlation-reduced photon index $\Gamma_{\rm cr}$.

\subsection{Distributions}
\label{dist}
\begin{figure}[H]
\includegraphics[width=7.0 cm]{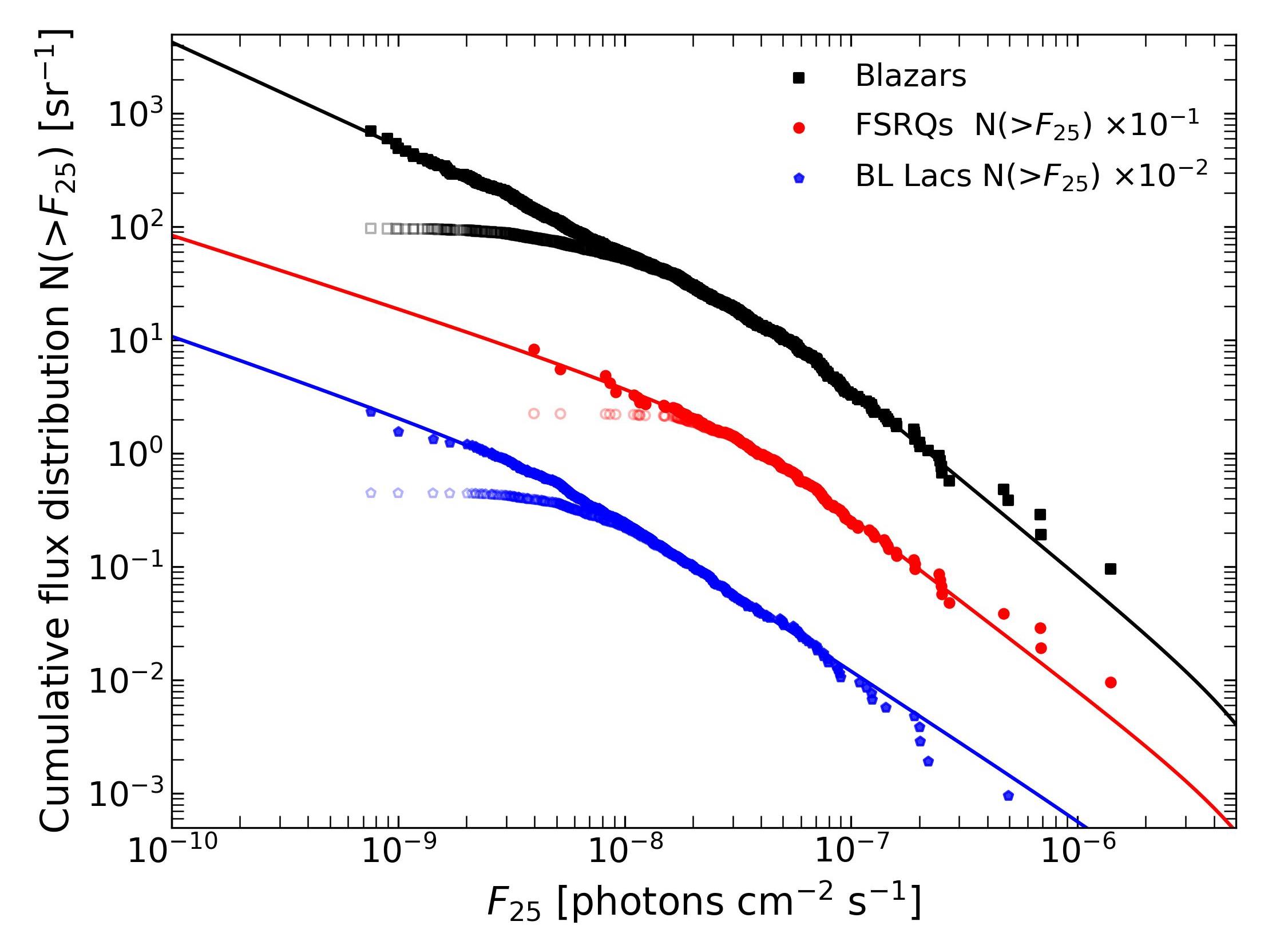}
\includegraphics[width=7.0 cm]{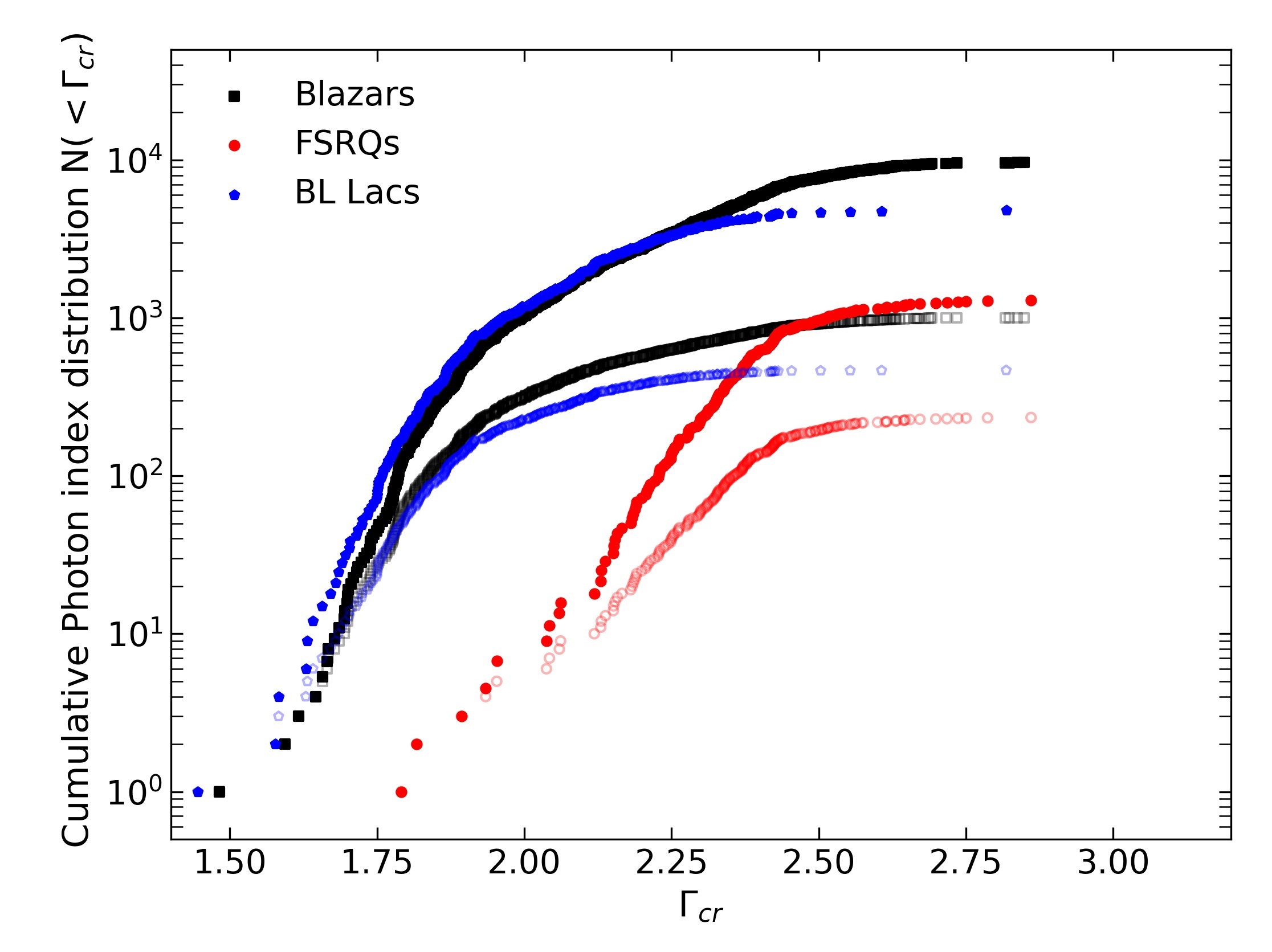}
	\caption{\textbf{Left}: The cumulative photon flux distribution (solid points), $\Psi(F_{25})$, corrected for the photon flux limit, and compared with the raw histogram of photon flux (open points), $N(>F_{25})$.
	The solid curves display the cumulative distributions of flux achieved by integrating the broken power-law distribution (Equation (\ref{eq:broken})) using the best-fit parameters listed in Table \ref{tab:fitpatameters}. \textbf{Right}: The cumulative photon index distribution $H(\Gamma_{\rm cr})$ vs. $\Gamma_{\rm cr}$ corrected for selection effects (solid points) and compared with the raw cumulative observed photon index (open points), $\Gamma$, $N(>\Gamma)$. } 
	\label{fig:cumdist}
\end{figure}

Upon acquiring the truncated data with independence in the $F_{25} - \Gamma_{\rm cr}$ plane, we apply the Lynden-Bell method, as developed by \citet{1971MNRAS.155...95L}, to derive the cumulative mono-variate distributions $\Psi(F_{25})$ and $H(\Gamma_{\rm cr})$.  This method provides a nonparametric and point-by-point description of the cumulative distributions, conventionally sorted in descending order by photon flux and ascending order by photon index (see also, e.g.,~\citet{2015ApJ...806...44P,2021ApJ...913..120Z}). Then, their distributions are represented as follows:
	\begin{equation}
	\label{associated}
	\Psi(F_{25,i})=\prod_{j=1}^{i}\left(1+\frac{1}{N_j}\right) \,\,\,\,{\rm and}\,\,\,\, H(\Gamma_{\rm cr, i})=\prod_{j=1}^{i}\left(1+\frac{1}{M_j}\right),
	\end{equation}
where $N_j$ represents the number of objects in the associated set of
object $i$ (with variables $F_{25,i}, \Gamma_{{\rm cr},i}$) consisting of sources with $F_{25,j}>F_{25,i}$ and $\Gamma_{{\rm cr},j}<\Gamma_{{\rm cr},{\rm max},i}$ (or $F_{{\rm min},j}<F_{25,i}$);
and $M_j$ denotes the number of objects in the associated set of
object $i$ comprising sources with $\Gamma_{{\rm cr},j}<\Gamma_{{\rm cr},i}$ and $F_{25,j}>F_{{\rm min},i}$ (or $\Gamma_{{\rm cr},{\rm max},j}>\Gamma_{{\rm cr},i}$). 
The cumulative distribution of flux and photon index values can also be expressed as integrals of the independent distributions $\psi(F_{25})$ and $\hat{h}(\Gamma_{\rm cr})$, as follows:
	\begin{equation}
	\Psi(F_{25})= \int_{F_{25}}^{\infty} \psi(F_{25}^{\prime}) dF_{25}^{\prime}\,\,\,\,{\rm and} \,\,\, H(\Gamma_{\rm cr})=\int_0^{\Gamma_{\rm cr}} \hat{h}(\Gamma_{\rm cr}^\prime)d\Gamma_{\rm cr}^\prime. 
	\label{cumdist}
	\end{equation}
Figure \ref{fig:cumdist} illustrates the cumulative distributions of photon flux $F_{25}$ and index $\Gamma_{\rm cr}$ corrected for the photon flux limit, alongside the corresponding raw cumulative distributions of $F_{25}$ and $\Gamma_{\rm cr}$, denoted as $N(>F_{25})$ and $N(<\Gamma_{\rm cr})$, respectively. 

The derivatives of the cumulative distributions yield the following differential distributions:
	\begin{equation}
	 \label{diffdist}
	\psi(F_{25})=-{d\Psi(F_{25})\over dF_{25}}, \,\,\,\, {\rm and} \,\,\,\, \hat{h}(\Gamma_{\rm cr}) =  \frac{dH(\Gamma_{\rm cr})}{d\Gamma_{cr}}.
	\end{equation}
Their distributions can be obtained directly by differentiation of the numerical histograms, as described in Appendix B of \cite{2015ApJ...806...44P}. However, a point-by-point derivative may result in noise, which can be mitigated by smoothing. 

\begin{figure}[H]
\includegraphics[width=7.0 cm]{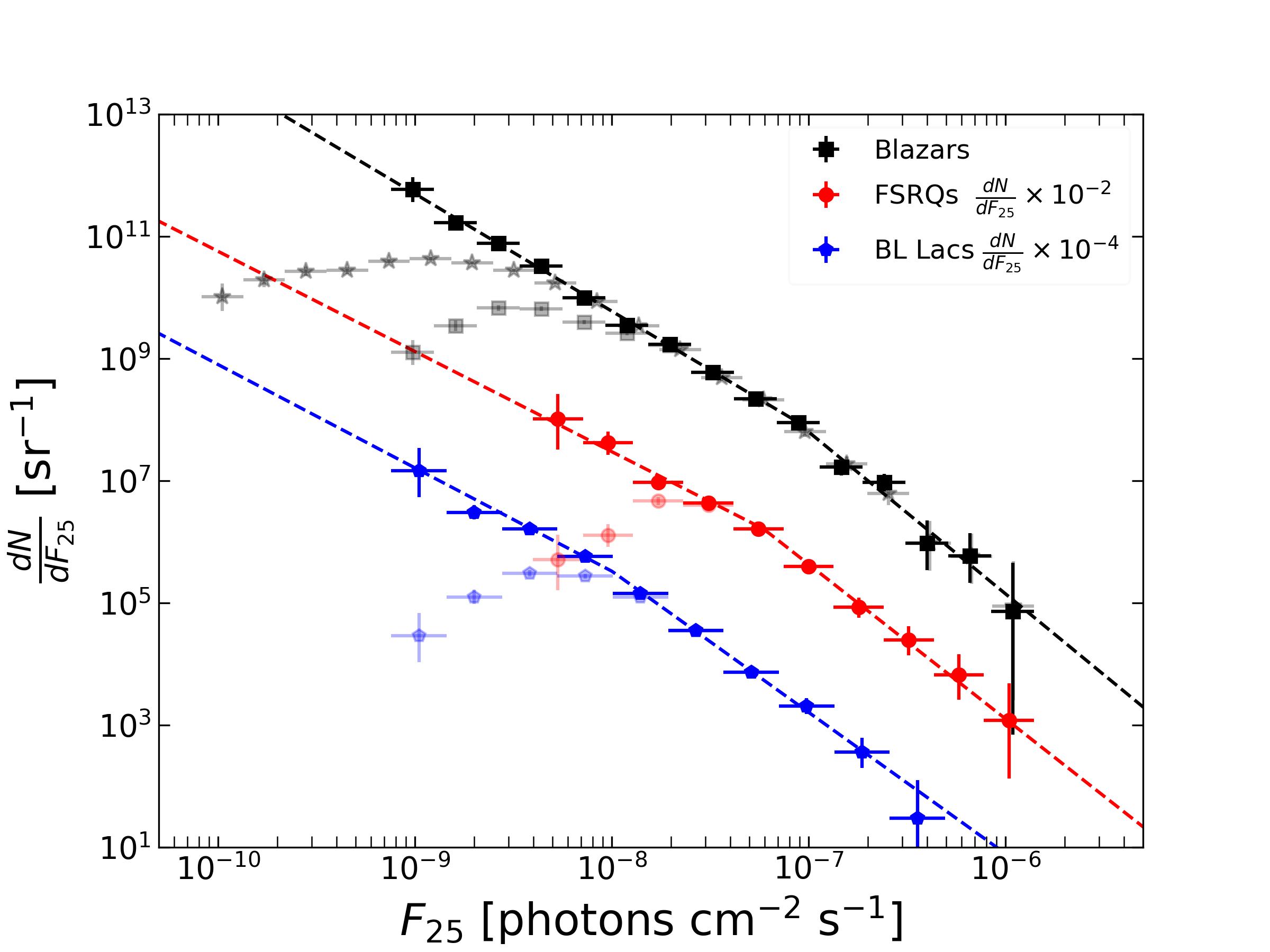}
\includegraphics[width=7.0 cm]{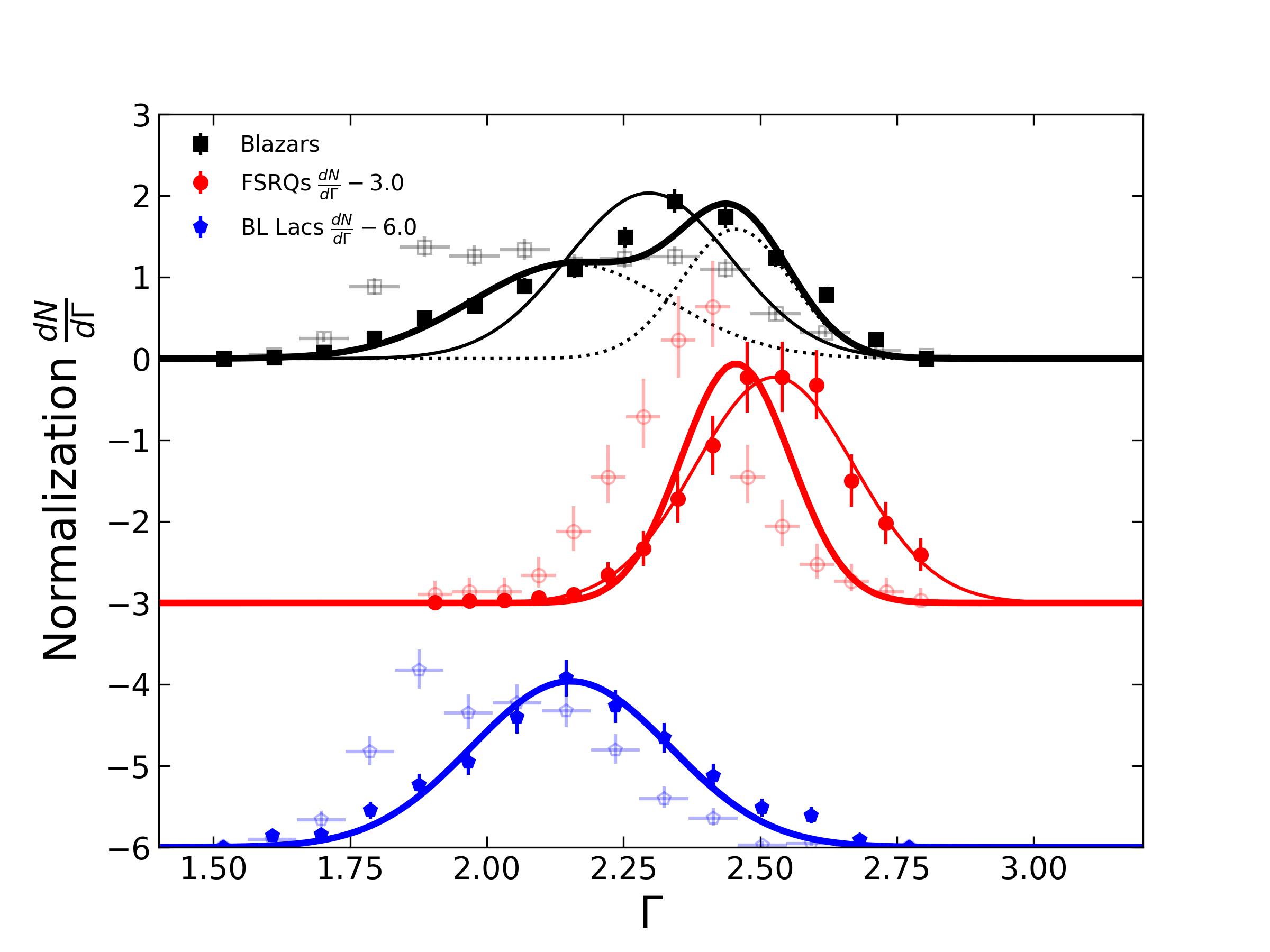}
	\caption{The differential intrinsic photon flux distribution (left) and photon index distribution (right) for blazars, FSRQs, and BL Lacs, separately.
     The $dN/dF_{25}$ values derive from the differentiation of the cumulative photon flux distribution for corrected and raw data, and the curves derive from Equation (\ref{eq:broken}) with the best-fit parameters. The intrinsic $dN/d\Gamma = h(\Gamma)$ values are shown as solid points, which derive from Equation (\ref{eq:hGam}), and the raw observed distributions are shown as open points. The Gaussian distribution form is used to fit $h(\Gamma)$ and show the best-fit curves. The raw distribution of observations with 2796~blazars is also shown in the left panel (the open stars).} 
	\label{fig:diffdist}
\end{figure}

For flux, employing interpolation smoothing provides the differential distributions shown by the points with error bars in the left panel of Figure \ref{fig:diffdist}. To facilitate comparison with the results of previous studies (e.g., \citep{2010ApJ...720..435A,2012ApJ...753...45S}), we assume that the intrinsic distribution of flux follows a broken power-law model:
\begin{eqnarray}
\psi(F_{25})= A \times 
 \left\{
\begin{array}{lcl}
 F_{25}^{-\beta_1}, &~~~& {\rm if}~~~ F_{25} \ge F_{b}; \\
F_{b}^{\beta_2-\beta_1} F_{25}^{-\beta_2}, &~~~& {\rm if}~~~F_{25} \le F_{b}\;,
\end{array}
\right.
\label{eq:broken}
\end{eqnarray}
where $A$ is the normalization and $F_{b}$ is the flux break. $\beta_1$ and $\beta_2$ are the power-law slopes above and below the break, respectively. 
Parameters obtained via Markov Chain Monte Carlo (MCMC) fitting of the cumulative distribution of photon flux are listed in Table  \ref{tab:fitpatameters}, computed using Equation (\ref{cumdist}). 
This method necessitates providing the likelihood function $L$, where we define the $\chi^2= -2{\rm ln} L$ and $\chi^2=\Sigma_i^n \frac{(Y_i-N_i)^2}{\sigma_i^2}$. Here, $Y_i$ represents $\Psi(F_{25,i})$ and $N_i$ represents $N(>F_{25,i})/10.38$ sr, indicating the number of objects above $F_{25,i}$ per steradian. 
The uncertainties associated with $N_i$ are typically treated as Poisson errors, represented by $\sigma= \sqrt{N_i}/10.38$ sr for $F_{25,i} > 3 \times 10^{-8}$ photons cm$^{-2}$ s$^{-1}$, and $\sigma= \frac{N_i}{N(>3 \times 10^{-8})}\sqrt{N(>3 \times 10^{-8})}$/10.38 sr otherwise. The truncation of data below $F_{25} < 3 \times 10^{-8}$ photons cm$^{-2}$ s$^{-1}$ is significant, warranting error scaling relative to $\Psi(F_{25}=3 \times 10^{-8})$. This error treatment method was initially introduced in \cite{2012ApJ...753...45S}, providing comprehensive details. Notably, the total sky coverage considered is 10.38 sr, corresponding to $|b| > 10^o$. The left panels of Figures \ref{fig:cumdist} and \ref{fig:diffdist} illustrate that these distributions are well characterized by a broken power-law form for both blazars and their subsets. 

Regarding the photon index, Equation (\ref{eq:hGam}) calculates the intrinsic distribution of the photon index, $h(\Gamma)$, using the parameterized distribution $\psi(F_{25})$ and numerical differentiation, resulting in $\hat{h}(\Gamma_{\rm cr})$. The right panel of Figure \ref{fig:diffdist} demonstrates the intrinsic distributions of the photon index and the raw observed distributions for blazars, FSRQs and BL Lacs. It should be noted that the best-fit value of the correlation parameter $\beta$ is utilized here.
A Gaussian model was applied to fit the intrinsic distribution of the photon index. Interestingly, the distribution for blazars deviates from a Gaussian distribution, while those for FSRQs and BL Lacs are well described. Notably, the intrinsic photon index values' distribution for blazars seems better represented by two Gaussian distributions, suggesting a non-Gaussian nature. 

The best-fit model of the intrinsic source count distribution is a broken power-law model for blazars, FSRQs, and BL Lacs, with the break points found at $F_b=8.39^{+1.16}_{-1.07} \times 10^{-8}$, $5.92^{+1.31}_{-1.02} \times 10^{-8}$, and $1.01^{+0.31}_{-0.39} \times 10^{-8}$ ph cm$^{-2}$ s$^{-1}$, respectively. The slopes above and below the break are $2.65^{+0.14}_{-0.12}$ and $1.92^{+0.01}_{-0.01}$ for blazars, $2.53^{+0.16}_{-0.11}$ and $1.64^{+0.09}_{-0.12}$ for FSRQs, and $2.31^{+0.05}_{-0.06}$ and $1.69^{+0.13}_{-0.45}$ for BL Lacs. 
The intrinsic distributions of the photon indices of FSRQs and BL Lacs are consistent with Gaussian distributions, with means of $2.53^{+0.02}_{-0.02}$ and $2.16^{+0.02}_{-0.02}$, and dispersions of $0.15^{+0.02}_{-0.02}$ and $0.18^{+0.01}_{-0.01}$, respectively. However, a Gaussian model with a mean and dispersion of $2.30^{+0.02}_{-0.02}$ and $0.15^{+0.01}_{-0.01}$ is not a good fit for the distribution of blazars. Based on the fact that each source in the blazar sample is either an FSRQ or a BL Lac, a simultaneous joint fit is performed,  constraining the means and widths of the two Gaussian distributions in the blazar sample to equal those in the respective individual BL Lac and FSRQ samples. The mean values of the two Gaussians are, respectively, $2.46^{+0.02}_{-0.02}$ and $2.15^{+0.02}_{-0.02}$, while the corresponding dispersions are $0.10^{+0.02}_{-0.02}$ and $0.18^{+0.01}_{-0.01}$. The results of individual and joint fits are shown in right panel of Figure~\ref{fig:diffdist}, represented by the thin and the thick solid lines. 
 The best-fit parameters are shown in the Table \ref{tab:fitpatameters}, indicating general agreement with the findings of Refs. \cite{2010ApJ...720..435A,2012ApJ...753...45S}. 

We note that $\beta_1 = 2.31^{+0.05}_{-0.06}$ for BL Lacs deviates from the expected value of 2.5 for a Euclidean distribution in the local universe. Additionally, the break flux $F_b$ in the source count distribution for BL Lacs is lower than that reported in \cite{2010ApJ...720..435A,2012ApJ...753...45S}. Abdo et al. \cite{2010ApJ...720..435A} proposed that this break is of cosmological origin, coinciding with changes in population evolution sign. BL Lac objects demonstrate positive cosmological evolution at high luminosities and negative evolution at low luminosities \cite{Ajello_2013}, potentially leading to a source count distribution following a three-component power law. In the left panels of  \linebreak Figures \ref{fig:cumdist} and \ref{fig:diffdist}, it can be seen that there appears to be a break flux at $F_b \sim 6.0 \times 10^{-8}$ ph cm$^{-2}$ s$^{-1}$.

\begin{table}[H]
\renewcommand{\arraystretch}{1.3}	
\caption{The best-fit parameters for Fermi-LAT blazars of photon flux and index intrinsic distribution.\label{tab:fitpatameters}}
\begin{adjustwidth}{-\extralength}{0cm}
\begin{tabularx}{\fulllength}{lC|CCCC|CC}
\noalign{\hrule height 1pt}
\textbf{Sample}	& \boldmath{\textbf{$\beta^{\text{a}}$}}	& \textbf{log10(\boldmath{$A^{\text{b}}$})}& \textbf{\boldmath{$\beta_1^{\text{c}}$}}& \textbf{\boldmath{$F_{b}^{\text{d}}$}}& \textbf{\boldmath{$\beta_2^{\text{e}}$}}& \textbf{\boldmath{$\mu^{\text{f}}$}}& \textbf{\boldmath{$\sigma^{\text{g}}$}}\\
\hline
Blazars (this work)	& $0.006^{+0.033}_{-0.033}$ &$-10.75^{+0.79}_{-0.93}$&  $2.65^{+0.14}_{-0.12}$ & $8.39^{+1.16}_{-1.07}$ & $1.92^{+0.01}_{-0.01}$ & $2.30^{+0.02}_{-0.02}$& $0.15^{+0.01}_{-0.01}$ \\
Blazars (EP-L) & $0.02 \pm 0.08$ &$-$ &$2.37\pm0.13$ & $7.0 \pm 0.2$ & $1.70 \pm 0.26$ &$2.41\pm0.13 $& $0.25 \pm 0.03$ \\
Blazars (MA) & $-$ & $-$& $2.48 \pm 0.13$ & $7.39 \pm 1.01$ & $1.57 \pm 0.09$ & $2.37\pm0.02 $& $0.28 \pm 0.01$ \\
\hline
\multirow{2}*{FSRQs (this work)} & \multirow{2}*{$-0.100^{+0.031}_{-0.038}$} & \multirow{2}*{$-10.08^{+0.77}_{-1.13}$}& \multirow{2}*{$2.53^{+0.16}_{-0.11}$} & \multirow{2}*{$5.92 ^{+1.39}_{-1.02}$} & \multirow{2}*{$1.64 ^{+0.09}_{-0.12}$ }& $2.53^{+0.02}_{-0.02}$& $0.15^{+0.02}_{-0.02}$ \\
\cline{7-8}
&&&&&& $2.46^{+0.02}_{-0.02}$&$0.10^{+0.02}_{-0.02}$\\
FSRQs (EP-L)&  $-0.11 \pm 0.06$ &$-$ & $2.22\pm0.09$ & $5.1 \pm 2.0$ & $1.62 \pm 0.46$ & $2.52\pm0.08 $& $0.17 \pm 0.02$ \\
FSRQs (MA)&  $-$ &$-$ &$2.41\pm0.16$ & $6.12 \pm 1.3$ & $0.70 \pm 0.30$ &$2.48\pm0.02 $& $0.18 \pm 0.01$ \\
\hline
\multirow{2}*{BL Lacs (this work)} & \multirow{2}*{$-0.017^{+0.037}_{-0.045}$} &\multirow{2}*{$-8.98^{+0.42}_{-0.33}$ }& \multirow{2}*{$2.31^{+0.05}_{-0.06}$ }& \multirow{2}*{$1.01 ^{+0.31}_{-0.39}$} & \multirow{2}*{$1.69 ^{+0.13}_{-0.45}$} & $2.16^{+0.02}_{-0.02}$& $0.18 ^{+0.01}_{-0.01}$ \\
\cline{7-8}
&&&&&& $2.15^{+0.02}_{-0.02}$& $0.18 ^{+0.01}_{-0.01}$ \\
BL Lacs (EP-L)&  $0.04 \pm 0.09$ &$-$ & $2.55\pm0.17$ & $6.5 \pm 0.5$ & $1.61 \pm 0.27$ & $2.13\pm0.13 $& $0.24 \pm 0.02$ \\
BL Lacs (MA)&  $-$ &$-$ & $2.74\pm0.03$ & $6.77 \pm 1.30$ & $1.72 \pm 0.14$ & $2.18\pm0.02 $& $0.23 \pm 0.01$ \\
\noalign{\hrule height 1pt}
\end{tabularx}
\end{adjustwidth}
	\noindent{\footnotesize{Notes. EP-L: the results of \cite{2012ApJ...753...45S}; MA: the results of \cite{2010ApJ...720..435A}.}
	\footnotesize{\textsuperscript{a} The correlation between photon index $\Gamma$ and flux $F_{25}$---see Equation (3) and Section \ref{sec3.1}.}
	\footnotesize{\textsuperscript{b} The intrinsic flux distribution $\psi(F_{25})$ at break flux $F_b$, in units of cm$^{2}$ s$^{1}$ sr$^{-1}$.}
	\footnotesize{\textsuperscript{c} The power-law slope of the intrinsic flux distribution $\psi(F_{25})$ at fluxes above the break in the distribution.}
	\footnotesize{\textsuperscript{b} The flux at which the power-law break in $\psi(F_{25})$ occurs, in units of $10^{-8}$ ph cm$^{-2}$ s$^{-1}$ ($0.1 \leq E \leq 100$ GeV).}}
	\footnotesize{\textsuperscript{e} The power-law slope of the intrinsic flux distribution $\psi(F_{25})$ at fluxes below the break in the distribution.}
	\footnotesize{\textsuperscript{f} The mean of the Gaussian fit to the intrinsic photon index distribution $h(\Gamma)$.}	\footnotesize{\textsuperscript{g} The 1$\sigma$ width of the Gaussian fit to the intrinsic photon index distribution $h(\Gamma)$. The first and second rows represent the results of individual and joint fits, respectively. Joint fitting entails constraining the widths of the two Gaussians in the blazar sample to equal those in the respective individual BL Lac and FSRQ samples.}	
\end{table}
\vspace{-12pt}

\subsection{Detection Efficiencies}
\begin{figure}[H]
\includegraphics[width=10.5 cm]{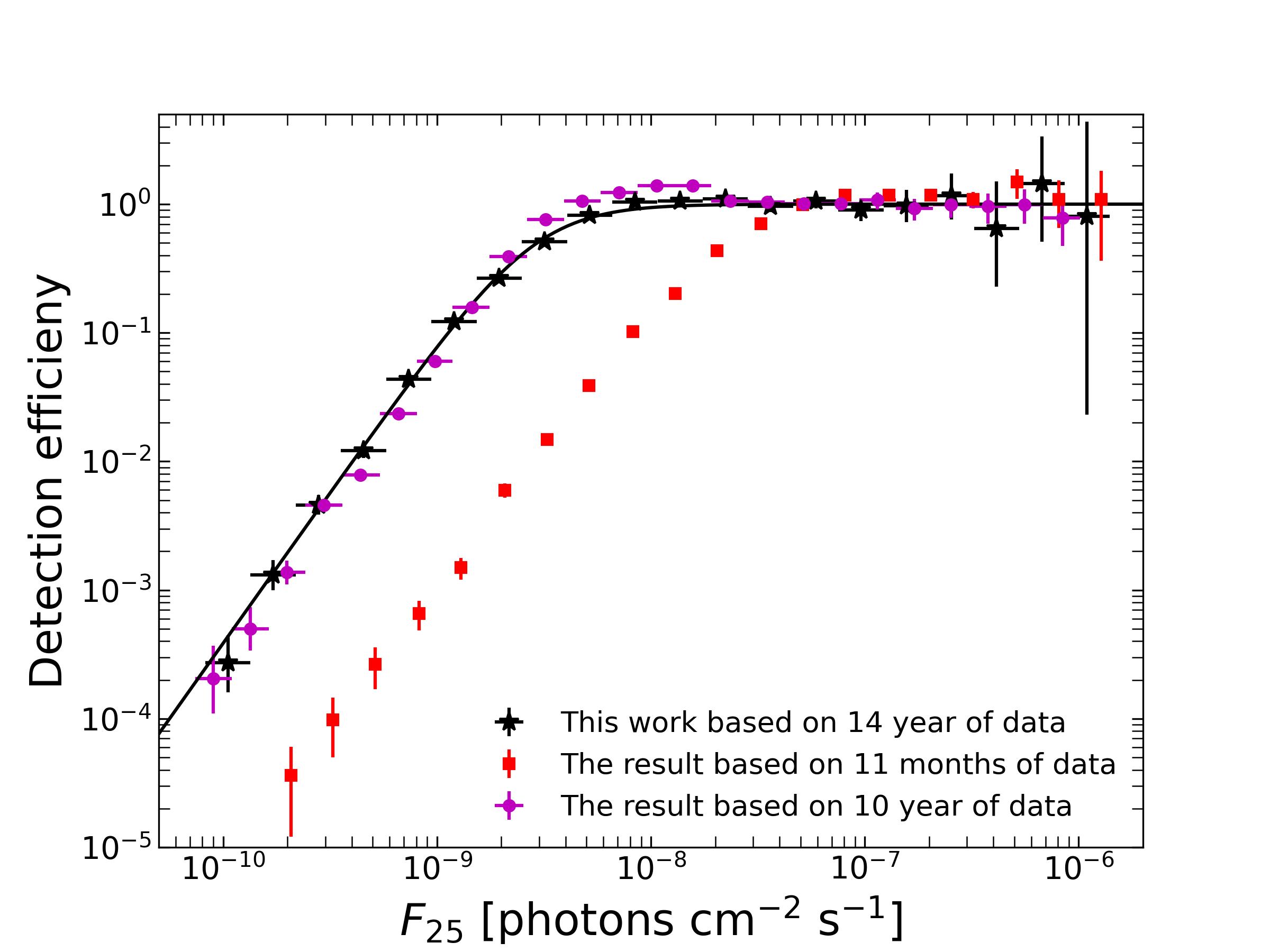}
	\caption{The detection efficiency $\omega (F_{25})$ for blazars, calculated by determining the ratio ({\bf Equation}~(\ref{eq:wf})) between the distribution of detected sources and the theoretical distribution. The red squares denote data obtained from~\cite{2010ApJ...720..435A}. The purple dots indicate the detection efficiency of the LAT high-latitude survey ($|b|>20^o$) with the simulated skies from \cite{2020ApJ...896....6M}, which is compared with our results.} 
	\label{fig:wf}
\end{figure}

Once the intrinsic differential photon flux distribution has been accurately determined, the detection efficiency of the sample can be calculated as follows:
\begin{equation}
    \omega(F_{25})=\frac{(dN/dF_{25})_{\rm obs}}{(dN/dF_{25})_{\rm th}}.
    \label{eq:wf}
\end{equation}
Here, we utilize the theoretical distribution $(dN/dF_{25})_{\rm th}$ rather than the smoothed differential distribution; specifically, we employ the best-fit distribution based on a broken power-law model. The observed distribution, $(dN/dF_{25})_{\rm obs}$, encompasses the entire sample from 14 years of Fermi-LAT operations. Due to significant incompleteness in the entire sample for FSRQs and BL Lac objects when analyzed separately, we focus solely on calculating the detection efficiency for all 2796 blazars. Both $(dN/dF_{25})_{\rm th}$ and $(dN/dF_{25})_{\rm obs}$ are illustrated in the left panel of Figure \ref{fig:diffdist}. The results are depicted in Figure \ref{fig:wf} as a black pentagram, along with the black curve described by $\omega(F)=(1+(F/F_0)^{-w_1})^{-w_2}$, where $F_0=2.88 \times 10^{-9}$ ph cm$^{-2}$ s$^{-1}$, $w_1=2.25$, and $w_2=1.04$, fitted using the least-squares method \cite{2019MNRAS.490..758Q}.
 Generally, detection efficiency is computed by providing lists of the simulated and detected sources for each efficiency simulation (see, e.g., \citep{2010ApJ...720..435A,2018ApJ...856..106D,2020ApJ...896....6M}).
Our method is also frequently employed for estimating detection efficiency, as seen in \cite{2014ApJ...780..161D,2019MNRAS.490..758Q}. It is noteworthy that our result is similar to that of \cite{2020ApJ...896....6M}, since the ratio $\sqrt{14}/\sqrt{10}=1.18$ closely approximates to~unity.

\subsection{Contribution to the EGB and IGRB}
Having derived the source count distribution and the detection efficiency, we now estimate the contribution of blazars to the EGB and the isotropic diffuse gamma-ray background (IGRB). The composition of the IGRB is dominated by undetected Fermi sources. We adopt a computational approach similar to that of Singal et al. \cite{2012ApJ...753...45S} and Zeng et al. \cite{2021ApJ...913..120Z}, which avoids the necessity for redshift evolution data or reliance on the luminosity function.
The differential background photon intensity from blazars can be determined as follows:
\begin{equation}
\frac{dN}{dEd\Omega}= \int^{F_{max}}_{F_{min}} dF_{25}
 \int^{\Gamma_{max}}_{\Gamma_{min}} d\Gamma
 \psi(F_{25})  h(\Gamma) f_\Gamma(E) 
 \left\{
 \begin{array}{lcl}
 (1-\omega(F_{25})) &~~~& {\rm for ~~~ IGRB}; \\
 1.0 &~~~& {\rm for ~~~ EGB},
 \end{array}
 \right.
 \label{eq:egb1}
\end{equation}
where $f_\Gamma(E)$ is the spectrum of sources. In the range of 100 MeV to 100 GeV, most blazars show a power-law spectrum, and the observed spectrum consists of a power law with exponential cutoff due to the optical depth $\tau(E,z)$, $f_\Gamma(E)=f_0\times E^{-\Gamma}e^{-\tau(E,z)}$, where $z$ is the redshift. Ignoring the effect of the optical depth $\tau(E,z)$ in the $E_1 = 100$ MeV to 100 GeV range, we can write $F_{25}=E_1^{(1-\Gamma)}(f_0/g(\Gamma))$, where
\begin{equation}
    g(\Gamma)=(\Gamma-1)/(1-10^{3(1-\Gamma)}).
\end{equation}
Substituting this into Equation (\ref{eq:egb1}), we obtain
\begin{equation}
\frac{dN}{dEd\Omega}=  P(E) \times \int^{F_{max}}_{F_{min}} dF_{25}
 \psi(F_{25}) F_{25}
 \left\{
 \begin{array}{lcl}
 (1-\omega(F_{25})) &~~~& {\rm for ~~ IGRB}; \\
 1.0 &~~~& {\rm for ~~ EGB},
 \end{array}
 \right.
 \label{eq:egb2}
\end{equation}
with 
\begin{equation}
P(E)=\frac{e^{-\tau(E,z)}}{E_1}
 \int^{\Gamma_{max}}_{\Gamma_{min}} d\Gamma
  h(\Gamma)g(\Gamma) \left(\frac{E}{E_1}\right)^{-\Gamma},
\end{equation}
where $P(E)$ is extracted from Equation (10) and encompasses the integration of the spectral index, which characterizes this particular spectral shape property.
We can obtain an analytic expression for $P(E)$ assuming a Gaussian distribution for $\Gamma$ with mean value of $\Gamma_0$ and dispersion $\sigma$, which gives
\begin{equation}
		P(E)=\frac{1}{\sqrt {2 \pi} \sigma}\frac{e^{-\tau(E,z)}}{E_1}\int_{-\infty}^\infty 
		e^{-(\Gamma-\Gamma_0)^2/2\sigma^2}(E/E_1)^{-\Gamma} g(\Gamma)d\Gamma.
\end{equation}
According to the analysis in \cite{2021ApJ...913..120Z}, this expression can be written as 
\begin{equation}
P(E)=[g(\Gamma^\prime)+g^{\prime\prime}(\Gamma^\prime)/4](E/E_1)^{-\Gamma_0}\frac{e^{\epsilon(E)-\tau(E,z)}}{E_1}\,\,\,\, {\rm with} \,\,\,\, 
\epsilon(E)={\sigma^2 [\ln(E/E_1)]^2 \over 2},  
\end{equation}
where $\Gamma^\prime(E)=\Gamma_0 - \sigma^2 \ln (E/E_1)$, 
 $g(\Gamma^\prime)=\sqrt{2}\sigma \times a/[1-e^{-ba)}]\,\,\,\, {\rm with}\,\,\,\, a=(\Gamma_0-1)/(\sqrt{2}\sigma)- (\sigma/\sqrt{2})\ln (E/E_1) \,\,\,\, {\rm and} \,\,\,\, b=9.8\sigma$, 
and the second derivative
\begin{equation}
g^{\prime\prime}(\Gamma^\prime)=-\sqrt{2}\sigma\times {be^{-ab}\over (1-e^{-ab})^3}[2(1-e^{-ab})-ba(1+e^{-ab})].
\end{equation}
Therefore, the integrated intensity photon energy above 100 MeV can be written as 
\begin{equation}
\frac{dN}{d\Omega}=\int_{100 \,\,\,{\rm MeV}} \frac{dN}{dEd\Omega}.
\label{Eq:integrate}
\end{equation}

\begin{figure}[H]
\includegraphics[width=7.0 cm]{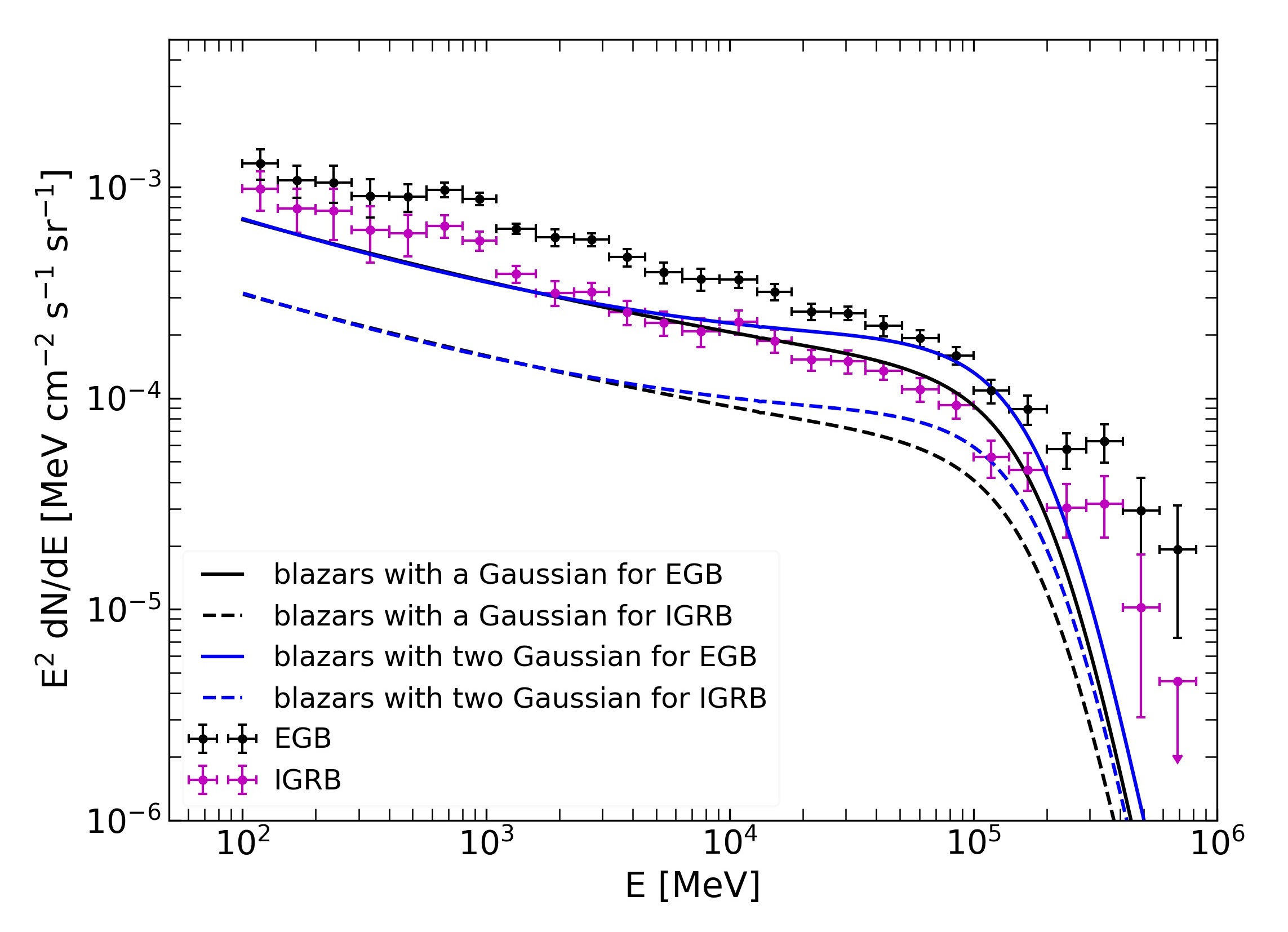}
\includegraphics[width=7.0 cm]{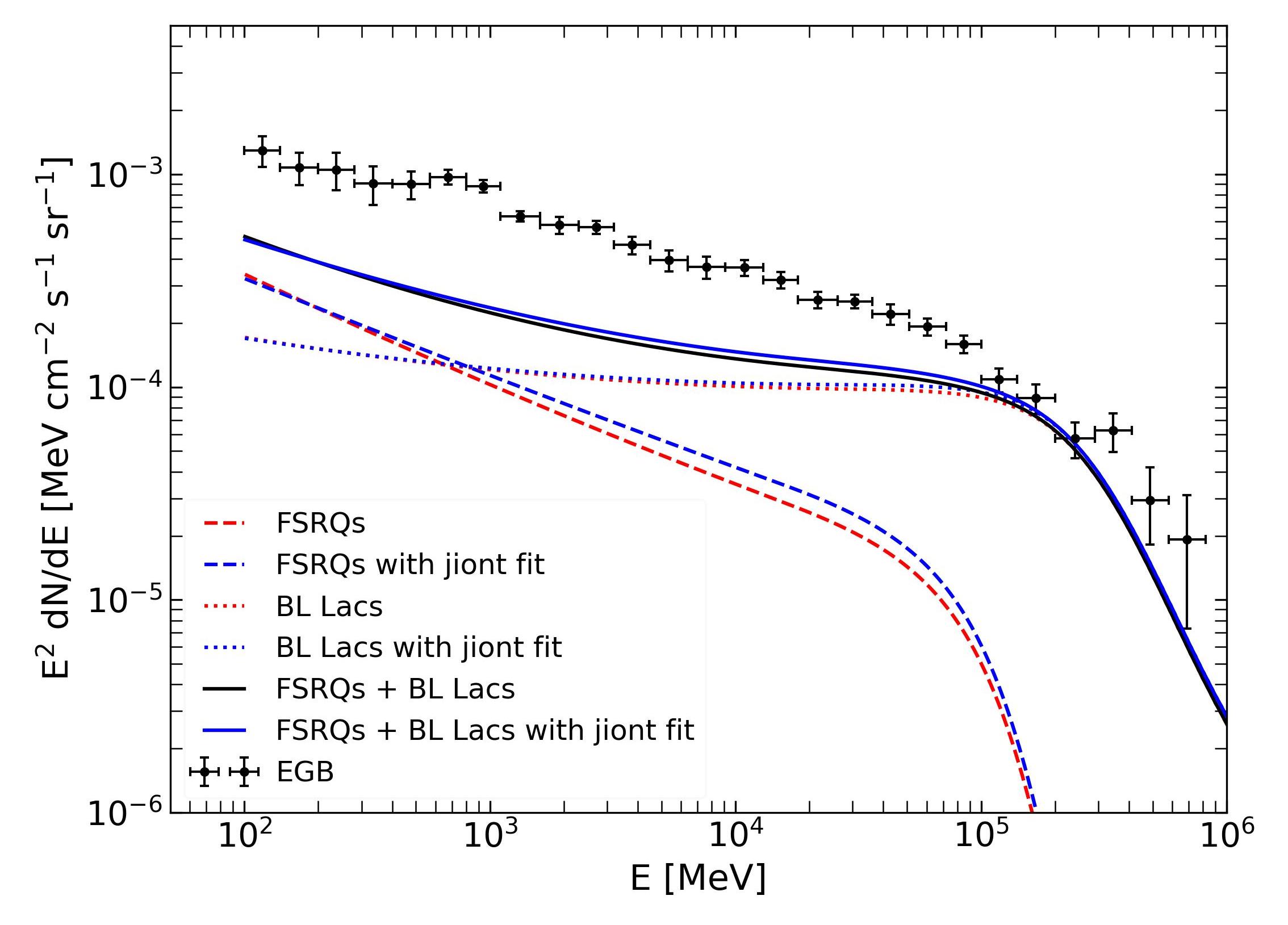}
	\caption{Left: The predicated EGB (solid line) and IGRB (dashed line) spectra for blazars, obtained by using the optical depth model of \cite{2010ApJ...712..238F} with $z=0.5$. The black and blue lines represent the intrinsic distribution of indexes used in the calculations as single Gaussian and double Gaussian, respectively. Right: The predicated EGB spectra for FSRQs with $z=1.0$ (dashed lines) and BL Lacs with $z=0.3$ ( dotted lines). The black solid line is the total EGB spectrum of the subset of blazars. The blue lines represent the results of the EGB spectrum calculated using the jointly fitted spectral index distribution. The observed EGB and IGRB data are taken from \cite{2015ApJ...799...86A}.} 
	\label{fig:EGB}
\end{figure}

Fermi-LAT has previously measured the intensities of the EGB and IGRB above 100 MeV as $1.13 \pm 0.07 \times 10^{-5}$ ph cm$^{-2}$ s$^{-1}$ sr$^{-1}$ and  $7.2 \pm 0.6 \times 10^{-6}$ ph cm$^{-2}$ s$^{-1}$ sr$^{-1}$, respectively, using 50 months of LAT data from 5 August 2008 and 6 October  2012 \cite{2015ApJ...799...86A}. According to Figure 8 in \cite{2015ApJ...810...14A}, the minimum photon flux detected from blazars by LAT over 48 months is approximately $ 5.0 \times 10^{-10}$ ph cm$^{-2}$ s$^{-1}$ with a spectral index $\Gamma \sim 1.5$. 
Thus, employing Equation (\ref{Eq:integrate}) with $F_{\rm min}=5.0 \times 10^{-10}$ ph cm$^{-2}$ s$^{-1}$, $F_{\rm max}=1.0 \times 10^{-5}$ ph cm$^{-2}$ s$^{-1}$, and the best-fit parameters in Table \ref{tab:fitpatameters}, we can estimate the contribution of blazars to the EGB and IGRB. The calculated contributions are $3.90 \times 10^{-6}$ ph cm$^{-2}$ s$^{-1}$ sr$^{-1}$ for the EGB and  $1.21 \times 10^{-6}$ ph cm$^{-2}$ s$^{-1}$ for the IGRB, representing 34.5 and 16.8 percent of the intensities observed by Fermi-LAT, respectively. 
The predicated EGB and IGRB spectra of blazars, using Equation (\ref{eq:egb2}) with the different intrinsic distributions of the indexes, are shown in the left panel of Figure~\ref{fig:EGB}. Extrapolating to zero flux, we find that blazars in total can produce a photon output of $8.27 \times 10^{-6}$ ph cm$^{-2}$ s$^{-1}$ sr$^{-1}$ to the EGB and $5.72 \times 10^{-6}$ ph cm$^{-2}$ s$^{-1}$ to the IGRB. 

The detection efficiencies for FSRQs and BL Lacs are not obtained in this work; we solely estimated their contribution to the EGB. Our results indicate the contribution from FSRQs to be $2.22 \times 10^{-6}$ ph cm$^{-2}$ s$^{-1}$ sr$^{-1}$, which accounts for 19.6 percent of Fermi's EGB intensity, and the contribution from BL Lacs to be $1.47 \times 10^{-6}$ ph cm$^{-2}$ s$^{-1}$ sr$^{-1}$, representing 13 percent of the Fermi EGB intensity. Consequently, their combined contribution amounts to 32.6 percent of the Fermi EGB intensity. The right panel of Figure \ref{fig:EGB} illustrates the predicated EGB spectra of FSRQs, with $z=1.0$, and BL Lacs, with $z=0.3$. When using Equation (\ref{Eq:integrate}) with $F_{\rm min}=0$ to assess their maximum contributions to the EGB, FSRQs contribute $2.36 \times 10^{-6}$ ph cm$^{-2}$ s$^{-1}$ sr$^{-1}$ and BL Lacs contribute $1.77 \times 10^{-6}$ ph cm$^{-2}$ s$^{-1}$ sr$^{-1}$. When the jointly fitted spectral index distribution is utilized, the intensity of the EGB increases slightly due to the harder spectrum. As shown in right panel of Figure \ref{fig:EGB}, the contribution from FSRQs exceeds that of BL Lacs at energies below approximately 600 MeV, whereas the contribution from BL Lacs dominates the EGB at energies around 600 MeV. 
	Notably, this boundary energy is higher than the value reported in \cite{2010ApJ...720..435A} (300 MeV). According to Equation (\ref{Eq:integrate}), with $E_{\text{min}} = 50$ GeV and $F_{\text{min}} = 0$, the total integrated flux from BL Lacs is found to be $1.98 \times 10^{-9}$ ph cm$^{-2}$ s$^{-1}$ sr$^{-1}$, which accounts for about 78\% of the EGB above 50 GeV.
	This estimate is lower than that reported in \cite{2019MNRAS.490..758Q}, at about 100\%, which was obtained through employing the gamma-ray luminosity function of BL Lacs, and lower than that reported in \cite{2016PhRvL.116o1105A}, at $84^{+16}_{-14}$ \%, which was obtained after the authors analyzed a sample from the second catalog of hard Fermi-LAT sources (2FHL) consisting of 90\% BL Lacs. Furthermore, the contribution of FSRQs to the EGB is approximately one-tenth that of BL Lacs, about $1.5 \times 10^{-10}$ ph cm$^{-2}$ s$^{-1}$ sr$^{-1}$; this disparity can be attributed to the larger spectral index and significant absorption effects at higher redshifts. 

It is noteworthy that here, we utilize the EBL model of \cite{2010ApJ...712..238F} to calculate $\tau(E,z)$, which requires a redshift parameter. However, we do not consider redshift evolution of sources but aim to select characteristic redshifts to represent the average evolutionary effects or assume no evolution with a constant redshift. Base on the redshift distribution of 3LAC \citep{2015ApJ...810...14A} and 4LAC \citep{2020ApJ...892..105A}, FSRQs exhibit a broad peak around $z=1$, while the predominant peak occurs at $z=0.3$ for BL Lacs. Therefore, we employ specific redshift values of 0.5, 1.0, and 0.3 to calculate $\tau(E,z)$ for the entire dataset, for FSRQs, and for BL Lacs, respectively. As depicted in Figure \ref{fig:EGB}, the predicted EGB spectrum exhibits an earlier cutoff compared to the observed data, suggesting that the characteristic redshift of blazars and BL Lacs may be lower than those aforementioned values.

\section{Summary and Discussion}\label{sec4}
In this study, we have employed the rigorous nonparametric methods of Efron--Petrosian and Lynden-Bell (EP-L) to calculate the source count distribution of blazars based on gamma-ray photon fluxes and spectral indexes from the Fermi-LAT 14-year AGN catalog with $TS \geq 25$ at galactic latitude $|b|>10^o$. 
To address the significant presence of blazars with uncertain classifications, we mitigated the substantial classification incompleteness in subclass samples by adjusting the Fermi-LAT detection threshold based on flux, thus establishing a sample with 90\% classification completeness for comprehensive analysis. 
We investigated the correlation between photon flux and spectral index values across three blazar samples---blazars as a whole, FSRQs, and BL Lacs---and derived their intrinsic distributions in order to estimate their contributions to the EGB. Below is a concise summary of our procedures and~results:
\begin{enumerate}
\item \textit{Flux--index relations:} Addressing the bias towards hard sources in the photon flux limit (Figure \ref{fig:fgGam}),  we applied the EP method to correct for this bias, resulting in a correlation-corrected spectral index $\Gamma_{cr}$. Our analysis revealed no significant correlation for all blazars ($\beta=0.006 \pm 0.033$), with only a weak correlation being observed for FSRQs  ($\beta=-0.100 ^{+0.031}_{-0.038}$) and BL Lacs  ($\beta=-0.017^{+0.037}_{-0.045}$) individually. 
\item \textit{Intrinsic distributions:} Employing the Lynden-Bell method, we derived cumulative distributions and subsequently obtained differential distributions through numerical derivation. The true intrinsic differential distributions of flux exhibited a broken power law for blazars, FSRQs, and BL Lacs, individually. The intrinsic photon index distributions were described well by Gaussian forms for FSRQs and BL Lacs individually; however, when considering the sample as a whole, double Gaussians provided a better fit. Table 1 summarizes the best-fit parameters for the intrinsic flux and photon index distributions, comparing them with the results of previous studies (i.e., \cite{2010ApJ...720..435A,2012ApJ...753...45S}). 

\item \textit{Detection efficiency:} The source count distribution at $|b|>10^o$, shown in the left panel of Figure \ref{fig:diffdist}, 
displayed a broken power law for $F_{25}> 1.0 \times 10^{-8}$ ph cm$^{-2}$ s$^{-1}$. Below this flux threshold, the observed distribution dropped rapidly due to the challenges in detecting fainter sources with Fermi-LAT. We calculated the detection efficiency by comparing observed sources with the theoretical distribution in photon flux.

\item \textit{EGB and IGRB:}
Through the nonparametric determination of intrinsic distributions of photon flux and the spectral index, along with considerations of the detection efficiency, we directly calculated the contributions of all Fermi-LAT blazars to the EGB's and IGRB's intensity and spectra. Our findings indicate that blazars up to the flux threshold of Fermi-LAT, i.e., $F_{\rm min}=5.0 \times 10^{-10}$ ph cm$^{-2}$ s$^{-1}$,  could explain 34.5\% of observed EGB photons and 32.6\% of observed IGRB photons. For FSRQs and BL Lacs, we  estimated their contributions to the EGB without considering the detection efficiency, revealing that FSRQs and BL Lacs could account for 19.6\% and 13\% of the observed EGB, respectively. 
\end{enumerate}



\vspace{6pt} 

\authorcontributions{Conceptualization, H.Z.; formal analysis, X.H. and Z.H; investigation, X.Y.; writing---original draft preparation, X.Y. and H.Z; writing---review and editing, H.Z. All authors have read and agreed to the published version of the manuscript.}

\funding{This work is supported by the National Natural Science Foundation of China No. 11921003.}

\dataavailability{Files used to generate the various graphs presented in this paper are
available on request.
} 

\conflictsofinterest{The authors declare no conflicts of interest.} 

\acknowledgments{We are very grateful to the referees for the expert suggestions and the careful
corrections, which have significantly improved the paper.
}

\begin{adjustwidth}{-\extralength}{0cm}
\printendnotes[custom] 

\reftitle{References}

\PublishersNote{}
\end{adjustwidth}
\end{document}